\documentclass[reprint,superscriptaddress,onecolumn,showkeys,amsmath,amssymb,aps,prf]{revtex4-2}
\usepackage{graphicx}
\usepackage{soul}
\usepackage{amsmath}
\usepackage{dcolumn}
\usepackage{bm}
\usepackage[dvipsnames]{xcolor}
\usepackage{caption}
\usepackage{subcaption}
\begin{document}
\title{Capillary lubrication of a spherical particle near a fluid interface}
\author{Aditya Jha}
\affiliation{Univ. Bordeaux, CNRS, LOMA, UMR 5798, F-33405 Talence, France.}
\author{Yacine Amarouchene}
\affiliation{Univ. Bordeaux, CNRS, LOMA, UMR 5798, F-33405 Talence, France.}
\author{Thomas Salez}
 \email{thomas.salez@cnrs.fr}
\affiliation{Univ. Bordeaux, CNRS, LOMA, UMR 5798, F-33405 Talence, France.}
\date{\today}
\begin{abstract}
The lubricated motion of an object near a deformable boundary presents striking subtleties arising from the coupling between the elasticity of the boundary and lubricated flow, including but not limited to the emergence of a lift force acting on the object despite the zero Reynolds number. In this study, we characterize the hydrodynamic forces and torques felt by a sphere translating in close proximity to a fluid interface, separating the viscous medium of the sphere's motion from an infinitely-more-viscous medium. We employ lubrication theory and perform a perturbation analysis in capillary compliance. The dominant response of the interface owing to surface tension results in a long-ranged interface deformation, which leads to a modification of the forces and torques with respect to the rigid reference case, that we characterise in details with scaling arguments and numerical integrations. 
\end{abstract}
\maketitle

\section*{\label{sec:intro}Introduction}
The dynamics of objects moving in viscous fluids has been studied both theoretically and experimentally for a long time~\cite{lamb1924hydrodynamics, Batchelor1967,jeffery1915steady,collins1955steady,dean1963slow,o1964slow}. Confining the viscous flow between an object and a rigid surface modifies the forces felt by the object~\cite{o1967slow,goldman1967slow,cooley1969slow,jeffrey1981slow}. Such a modification is involved in vastly different phenomena ranging from the mechanics of joints~\cite{hou1992analysis,hlavavcek1993role}, to the movement of cells in capillaries~\cite{abkarian2002tank}, and the dynamics of suspensions~\cite{happel1983low,batchelor1970stress,batchelor1971stress,batchelor1976brownian,batchelor1977effect}. 

Recent research has provided evidence of boundary elasticity further modifying the lubricated dynamics of an object~\cite{leroy2011hydrodynamic,leroy2012hydrodynamic}. Further standardization of the measurement process has led to the design of contactless probes for rheology~\cite{garcia2016micro,Basoli2018}. The coupling of boundary elasticity and lubrication flow, collectively termed as soft lubrication, predicts the emergence of lift forces exerted on particles translating parallel to soft boundaries~\cite{sekimoto1993mechanism,Beaucourt2004,skotheim2005soft,Weekley2006,urzay2007elastohydrodynamic,Snoeijer2013,salez2015elastohydrodynamics,Bouchet2015,Essink2021,bertin2022soft,Bureau2023,rallabandi2024fluid}. Such lift forces are associated with the symmetry breaking arising out of the deformability. Since the latter is crucial to the generated force, the nature of the bounding wall has been further explored by examining the influence of slip~\cite{rinehart2020lift}, and viscoelasticity~\cite{Pandey2016,Kargar2021}. A reversal of the nature of the lift force from repulsive to attractive has also been predicted for viscoelastic settings~\cite{hu2023effect}. Other studies have explored the complex modifications induced by including inertial effects~\cite{clarke2011elastohydrodynamics} and compressibility~\cite{Balmforth2010}. On the experimental front, dedicated research has verified the presence of these lift forces on various substrates~\cite{Saintyves2016,Davies2018,Rallabandi2018,Vialar2019,zhang2020direct}. 

In biology, where cells and tissues are extremely soft, and/or at small scales in soft matter, the interfacial capillary stress at the boundary dominates over bulk elasticity. By employing a classical Stokeslet-like response of the flow near a fluid interface, it has been shown that a rectified flow may be generated owing to the tension of the boundary~\cite{aderogba1978action,nezamipour2021flow}. On the other hand, finite-size effects were addressed~\cite{lee1979motion,lee1980motion,lee1982motion,geller1986creeping} in the regime of a large gap between the object and the fluid interface, predicting counter-intuitive behaviors unique to capillarity. The results of these studies have been useful in analyzing the movement of microorganisms near a fluid interface~\cite{trouilloud2008soft,lopez2014dynamics}, as well as the formation of floating biofilms~\cite{desai2020biofilms}. Recently, capillary-lubrication studies~\cite{jha2023capillary,jha2024capillary} have characterized the dynamics of an infinite cylinder near a fluid interface, highlighting the influence of the viscosity contrast and thickness ratio between the two fluid layers, and leading to large variability in the forces generated as opposed to elastic interfaces. 

While previous research has highlighted the importance of understanding lubricated motion near a fluid interface, the characterization of the dynamics of a particle moving with multiple degrees of freedom near a fluid interface remains to be explored. In this article, we explore in detail the translational motion of a sphere moving in close proximity to an infinitely viscous but deformable sublayer. In the small-deformation limit, we calculate the forces and torques generated on the sphere during the motion. Due to a symmetry between translational and rotational motions in soft lubrication~\cite{bertin2022soft}, our work immediately generalizes to the case where rotation would be added. The remainder of the article is organized as follows. We start by describing the capillary-lubrication framework, before presenting the theoretical methodology for obtaining the different fields using perturbation analysis at small deformations of the fluid interface. We then discuss the implications of the interfacial deformation, and the competition between gravity and capillarity, on the forces and torques generated on the particle. Limiting expressions are derived for the capillary-dominated and gravity-dominated responses. While the former case is novel, the latter is reminiscent of a Winkler solid.

\section{\label{sec:capillary_lubrication_theory}Capillary-lubrication theory}
\begin{figure}[h]
\begin{center}
\includegraphics[width=8cm]{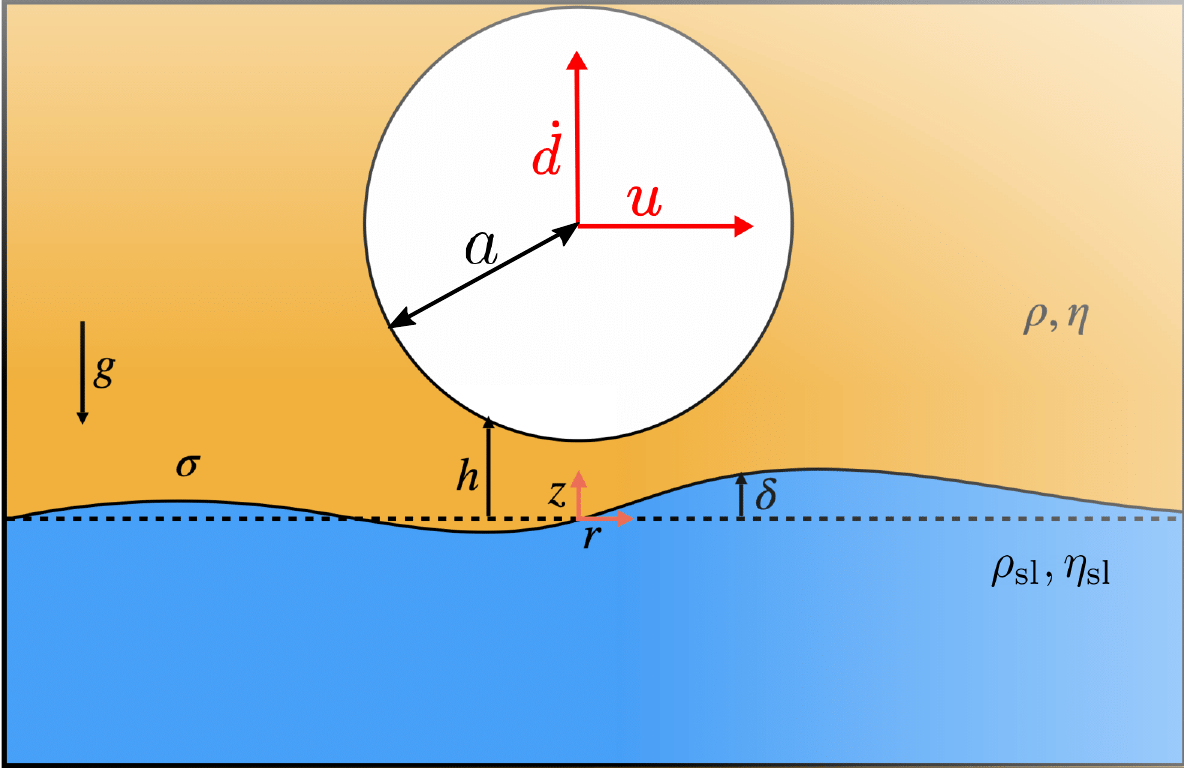}
\caption{Schematic of the system. A sphere of radius $a$ immersed in a viscous fluid of viscosity $\eta$ and density $\rho$ moves near a fluid interface. The undeformed gap profile is noted $h(r,t)$, with $r$ the horizontal radial coordinate and $t$ the time. The origin of coordinates is located at the undeformed fluid interface ($z = 0$) in line with the center of mass of the sphere ($r = 0$). The interface separates the top fluid from a secondary fluid of viscosity $\eta_{\textrm{sl}}$, with $\eta_{\textrm{sl}}\gg \eta$, and density $\rho_{\textrm{sl}}$ at the bottom, \textit{i.e.} $\rho_{\textrm{sl}}>\rho$. The sphere has prescribed horizontal velocity $u$ and vertical velocity $\dot{d}$, where $d=h(0,t)$ denotes the instantaneous distance between the sphere bottom and the undeformed fluid interface. The interface deflection field is denoted as $\delta(r,t)$, the acceleration of gravity is denoted as $g$, and the surface tension is denoted as $\sigma$.}
\label{fig:1}
\end{center}
\end{figure}

We consider a sphere of radius $a$ translating with a prescribed time-dependent horizontal velocity $\vec{u}=u(t)\vec{e}_x$ near a fluid interface, as shown in Fig.~\ref{fig:1}. The interface is characterized by its surface tension $\sigma$, and separates two incompressible Newtonian viscous liquids with dynamic shear viscosities $\eta$ and $\eta_{\textrm{sl}}$, as well as densities $\rho$ and $\rho_{\textrm{sl}}$ (with $\rho<\rho_{\textrm{sl}}$). The acceleration of gravity is denoted by $g$. The gap profile between the sphere and the undeformed fluid interface is denoted by $h(r,t)$, which depends on the horizontal radial coordinate $r$ and time $t$. The $x$-direction oriented along $\vec{e}_x$ corresponds to the horizontal angular coordinate $\theta=0$. The time-dependent distance between the bottom of the sphere surface and the undeformed interface is denoted by $d(t)$, with $d>0$. Its temporal derivative $\dot{d}(t)$ denotes the vertical velocity of the sphere. We focus on the case where the bottom layer is extremely viscous compared to the top layer, \textit{i.e.} $\eta_{\textrm{sl}}\gg\eta$, and is infinitely thick compared to the gap between the sphere and the undeformed interface. These conditions allow us to assume that the tangential velocity at the fluid interface vanishes because the shear stress from the top layer is insufficient to generate a flow in the lower layer. 

We neglect fluid inertia and assume that the typical gap between the sphere and the interface is much smaller than the typical horizontal length scale, allowing us to invoke lubrication theory~\cite{Reynolds1886,Oron1997}. Introducing the excess pressure field $p$, in the top layer, with respect to the hydrostatic contribution, and the horizontal velocity field $\vec{v}$ of the fluid in the gap, the incompressible Stokes equations thus read at leading lubrication order:
\begin{align}
\frac{\partial p}{\partial z} &= 0~,\label{eq:stokes1}\\
\nabla p&= \eta\frac{\partial^2 \vec{v}}{\partial z^2}~, \label{eq:stokes2}
\end{align}
where $\nabla$ denotes the gradient in the horizontal plane and where $z$ is the vertical coordinate.
In the limit of a small gap, the thickness profile of the fluid between the sphere and the undeformed fluid interface can be approximated by its parabolic expansion, leading to:
\begin{align} 
h(r,t) \simeq d(t)+\frac{r^{2}}{2 a}~.\label{eq:h} 
\end{align}
The no-slip boundary condition is imposed at both the sphere's surface and the fluid interface. Hence, in the frame of the moving sphere, at $z = h$, one has:
\begin{align}
\vec{v} = \vec{0}~, \label{eq:u_bndcn1}
\end{align} 
and at the interface, \textit{i.e.} $z = \delta$, one has:
\begin{align}
\vec{v} = -\vec{u} =-u\vec{e}_x~.  \label{eq:u_bndcn2}
\end{align}   
Given the boundary conditions above, we can easily write the horizontal velocity profile in the gap between the sphere and the interface, as: 
\begin{align}
\vec{v} = \frac{\vec{\nabla} p}{2\eta}(z-h)(z-\delta)+\vec{u}\,\frac{z-h}{h-\delta}~.\label{eq:gen_flow_profile}
\end{align}
The conservation of the fluid volume in the gap allows for the derivation of the Reynolds thin-film equation for the system, which reads: 
\begin{align}
\frac{\partial}{\partial t}(h-\delta) = \vec{\nabla}\cdot\left[\frac{\nabla p}{12\eta}(h-\delta)^3+\frac{\vec{u}}{2}(h-\delta)\right]~.\label{eq:dim_thin_film}
\end{align}
The deflection of the fluid interface is controlled by the Laplace pressure jump. Thus, at $z = \delta$, one has:  
\begin{align}
p& \simeq\sigma \nabla^2\delta+g\delta(\rho-\rho_{\textrm{sl}})~, \label{eq:laplace_dimensional}
\end{align}
where we have assumed a small deformation of the interface and linearized the curvature. 

Let us now non-dimensionalize the equations through: 
\begin{align*}
d(t)& = d^*D(T)~,   &h(r,t)& = d^*H(R,T)~,   & \vec{r} & =l\vec{R}~,  &  z & =d^*Z~, & t& = \frac{l}{c}T~,\\
\vec{v}(\vec{r},z,t)& =c\vec{V}(\vec{R},Z,T)~, &\vec{u}(t)& =cU(T)\vec{e}_x~, & p(\vec{r},t) &=\frac{\eta c l}{{d^*}^2}P(\vec{R},T)~, & \delta(\vec{r},t) & = d^*\Delta(\vec{R},T)~,
\end{align*}
where $c$ and $d^*$ represent characteristic in-plane velocity and gap-thickness scales, respectively, with $l = \sqrt{2ad^*}$ denoting the characteristic hydrodynamic radius. 

At this point, we discuss the assumption of an infinitely-viscous bottom layer and the resulting vanishing of the velocity, which has been previously addressed in the literature~\cite{yiantsios1990buoyancy}. Since the bottom fluid layer is considered to be much thicker than the upper one, the flow generated in the former evolves over a much larger length scale, typically given by $a$. As a consequence, by adapting the thin-bilayer-film normal stress balance (see Eq.~(A4) in~\cite{bertin2021capillary}) to the current case of a thick bottom layer, we find that the dimensionless normal viscous stress at the fluid interface has a $(\eta_{\textrm{sl}}/\eta)(d^*/l)^4$ prefactor. We assume this numerical prefactor to be very small in our study, due to the vanishing of $d^*/l$, despite the large viscosity ratio -- thus the non-inclusion of bottom-layer stress terms in Eq.~(\ref{eq:laplace_dimensional}).

In dimensionless terms, the undeformed thickness profile and Reynolds equation become:
\begin{align}
H(R,T) \simeq D(T)+R^{2}~, \label{eq:H} 
\end{align}
and:
\begin{align}
\frac{\partial}{\partial T}(H-\Delta) = \vec{\nabla}\cdot\left[\frac{\nabla P}{12}(H-\Delta)^3+\frac{\vec{U}}{2}(H-\Delta)\right]~.\label{eq:nondim_thin_film}
\end{align}
Besides, the deflection field $\Delta$ is related to the excess pressure field $P$ by the dimensionless version of the Laplace equation, which reads: 
\begin{align}
\nabla^2\Delta-\textrm{Bo}\Delta=\kappa P~, \label{eq:nondim_laplace}
\end{align}
where $\textrm{Bo} = (l/l_\textrm{c})^2$ denotes the Bond number, $l_\textrm{c} = \sqrt{\sigma/[g(\rho_{\textrm{sl}}-\rho)]}$ denotes the capillary length, and where we have introduced the capillary compliance of the interface, denoted by: 
\begin{align}
\kappa = \frac{\eta cl^3}{\sigma {d^*}^3}~. \label{eq:kappa}
\end{align}

\section{Perturbation analysis}
As in previous studies regarding soft lubrication~\cite{sekimoto1993mechanism,skotheim2005soft,urzay2007elastohydrodynamic,Snoeijer2013,salez2015elastohydrodynamics,Essink2021,bertin2022soft}, we perform an asymptotic expansion of the unknown fields at first order in dimensionless capillary compliance $\kappa$, as:
\begin{align}
\Delta &\simeq \kappa \Delta_{1}+O(\kappa^2)~, \\
P &\simeq P_{0}+\kappa P_{1}+O(\kappa^2)~, 
\end{align}
where $\kappa^i\Delta_i$ is the $i$-th order contribution to the deflection of the interface, and $\kappa^iP_i$ is the $i$-th order contribution to the excess pressure field.

\subsection{Zeroth-order pressure}
At zeroth order in $\kappa$, Eq.~(\ref{eq:nondim_thin_film}) reduces to: 
\begin{align}
\dot{D} = \nabla.\left(\frac{ \vec{\nabla} P_{0}}{12}H^3+\vec{U}\frac{H}{2}\right)\ . \label{eq:thin_film_0th}
\end{align}
This equation is identical to the one for a perfectly rigid, flat and no-slip boundary. Since the zeroth-order pressure field results from linear terms in velocity, we decompose it azimuthally, as:
\begin{align}
P_0(\vec{R},T) = P_{00}(R,T)+P_{01}(R,T)\cos\theta~, 
\end{align}
with $R$ and $\theta$ the horizontal polar coordinates of $\vec{R}$.
Assuming a vanishing pressure field at large $R$, and that it must remain finite and single-valued at $R=0$, the equations can be solved to give the zeroth-order components of the excess pressure field, as~\cite{o1967slow}: 
\begin{align}
P_{00} &= -\frac{3\dot{D}}{2(D+R^2)^2}~,\\
P_{01} &= \frac{6UR}{5(D+R^2)^2}~. \label{eq:zeroth_order_pressure}
\end{align}

\subsection{Interface deflection}
We now decompose the first-order deflection of the interface azimuthally, as: 
\begin{align}
\Delta_1(\vec{R},T) = \Delta_{10}(R,T)+\Delta_{11}(R,T)\cos\theta~.
\end{align}
Writing Eq.~(\ref{eq:nondim_laplace}) at first order in $\kappa$, one has:
\begin{align}
\frac{1}{R}\frac{\partial}{\partial R}\left(R\frac{\partial \Delta_{10}}{\partial R}\right)-\textrm{Bo}\Delta_{10}=P_{00}~, \label{eq:delta_10_diff_eqn}
\end{align}
and:
\begin{align}
\frac{1}{R}\frac{\partial}{\partial R}\left(R\frac{\partial \Delta_{11}}{\partial R}\right)-\frac{\Delta_{11}}{R^2}-\textrm{Bo}\Delta_{11}=P_{01}~, \label{eq:delta_11_diff_eqn}
\end{align}
with the boundary conditions $\Delta_{1i}\rightarrow0$ as $R\rightarrow\infty$, and finite and single-valued $\Delta_{1i}$ at $R=0$. The solutions of the above equations can be written in the most general form as: 
\begin{align}
\Delta_{10}(R) &= -I_0(\sqrt{\textrm{Bo}}R)\int_R^\infty K_0(\sqrt{\textrm{Bo}}\xi)\xi P_{00}(\xi)\textrm{d}\xi-K_0(\sqrt{\textrm{Bo}}R)\int_0^R I_0(\sqrt{\textrm{Bo}}\xi)\xi P_{00}(\xi)\textrm{d}\xi~, \label{eq:delta10_soln} \\
\Delta_{11}(R) &= -I_1(\sqrt{\textrm{Bo}}R)\int_R^\infty K_1(\sqrt{\textrm{Bo}}\xi)\xi P_{01}(\xi)\textrm{d}\xi-K_1(\sqrt{\textrm{Bo}}R)\int_0^R I_1(\sqrt{\textrm{Bo}}\xi)\xi P_{01}(\xi)\textrm{d}\xi~, \label{eq:delta11_soln}
\end{align}
where $I_j$ and $K_j$ denote the $j$th-order modified Bessel functions of the first and second kinds, respectively. 

To understand the parametric influence of the Bond number $\textrm{Bo}$, we explore the behaviors of the interface deflection for both small and large $\textrm{Bo}$ values. For vanishing $\textrm{Bo}$, the anisotropic deflection component $\Delta_{11}$ reaches a limiting behavior given by the following expression: 
\begin{align}
\Delta_{11} \simeq -\frac{3U}{10}\frac{\ln\left(1+\frac{R^2}{D}\right)}{R}~. \label{eq:delta_11_lim}
\end{align}
In contrast, the isotropic deflection component $\Delta_{10}$ does not show any limiting behavior at vanishing $\textrm{Bo}$, and a reduction in $\textrm{Bo}$ leads to an unbounded increase in $\Delta_{10}$. To understand the behavior of $\Delta_{10}$ as $\textrm{Bo}\rightarrow0$, we take an asymptotic approach previously used in problems relating to capillary deformations~\cite{james1974meniscus,lo1983meniscus,dupr2015shape}. The vanishing of $\textrm{Bo}$ allows for a scale separation in the radial coordinate $R$ into: i) an inner region controlled by surface tension; and ii) an outer region, where gravity is present. In the inner region ($R\ll1/\sqrt{\textrm{Bo}}$), the interface deformation denoted by $\Delta_{10}^{\textrm{i}}$ satisfies: 
\begin{align}
\frac{1}{R}\frac{\partial}{\partial R}\left(R\frac{\partial \Delta_{10}^{\textrm{i}}}{\partial R}\right)=P_{00}~. \label{eq:delta_10_inner_diff_eqn}
\end{align}
The general solution of the above equation reads: 
\begin{align}
\Delta_{10}^{\textrm{i}} = \mathcal{A}-\frac{3\dot{D}}{8D}\ln(D+R^2)~, \label{eq:delta_10_inner_soln}
\end{align}
where $\mathcal{A}$ is an integration constant. The far-field behavior of the inner solution reads: 
\begin{align}
\Delta_{10}^{\textrm{i}} \sim \mathcal{A}-\frac{3\dot{D}}{4D}\ln(R) ~. \label{eq:delta_10_inner_soln_limit}
\end{align}
In the outer region ($R \gg 1/\sqrt{\textrm{Bo}}$), gravity matters but there is no lubrication pressure, which leads to the following governing equation for the interface deformation denoted by $\Delta_{10}^{\textrm{o}}$: 
\begin{align}
\frac{1}{R}\frac{\partial}{\partial R}\left(R\frac{\partial \Delta_{10}^{\textrm{o}}}{\partial R}\right)-\textrm{Bo}\Delta_{10}^{\textrm{o}}=0~. \label{eq:delta_10_outer_diff_eqn}
\end{align}
The latter equation is solved, along with the condition that the deflection vanishes at infinity, leading to: 
\begin{align}
\Delta_{10}^{\textrm{o}} = \mathcal{B}K_0(\sqrt{\textrm{Bo}}R), \label{eq:delta_10_outer_soln}
\end{align}
where $\mathcal{B}$ is an integration constant. The near-field behavior of the outer solution reads: 
\begin{align}
\Delta_{10}^{\textrm{o}} \sim -\mathcal{B}\left(\gamma-\ln 2+\frac{\ln \textrm{Bo}}{2}+\ln R\right)~, \label{eq:delta_10_outer_soln_limit}
\end{align}
where $\gamma$ is Euler's constant. Matching Eq.~(\ref{eq:delta_10_inner_soln_limit}) with Eq.~(\ref{eq:delta_10_outer_soln_limit}) leads to: 
\begin{align}
\mathcal{A} &= -\frac{3\dot{D}}{4D}\left(\gamma-\ln 2+\frac{\ln \textrm{Bo}}{2}\right)~, \\
\mathcal{B} &= \frac{3\dot{D}}{4D}~.
\end{align}
Substituting these constants in Eqs.~(\ref{eq:delta_10_inner_soln}) and (\ref{eq:delta_10_outer_soln}), we find the matched asymptotes of the interface deflection at small $\textrm{Bo}$. The interface deflection can then be approximated by the matched crossover expression: 
\begin{align}
\Delta_{10}|_{\textrm{Bo}\rightarrow0} \approx \frac{3\dot{D}}{4D}\left[K_0(\sqrt{\textrm{Bo}}R)+\frac12\ln\left(\frac{R^2}{D+R^2}\right)\right]~.\label{eq:delta_10_inner_soln_matched}
\end{align}
\begin{figure}[h]
\begin{center}
\includegraphics[width=10cm]{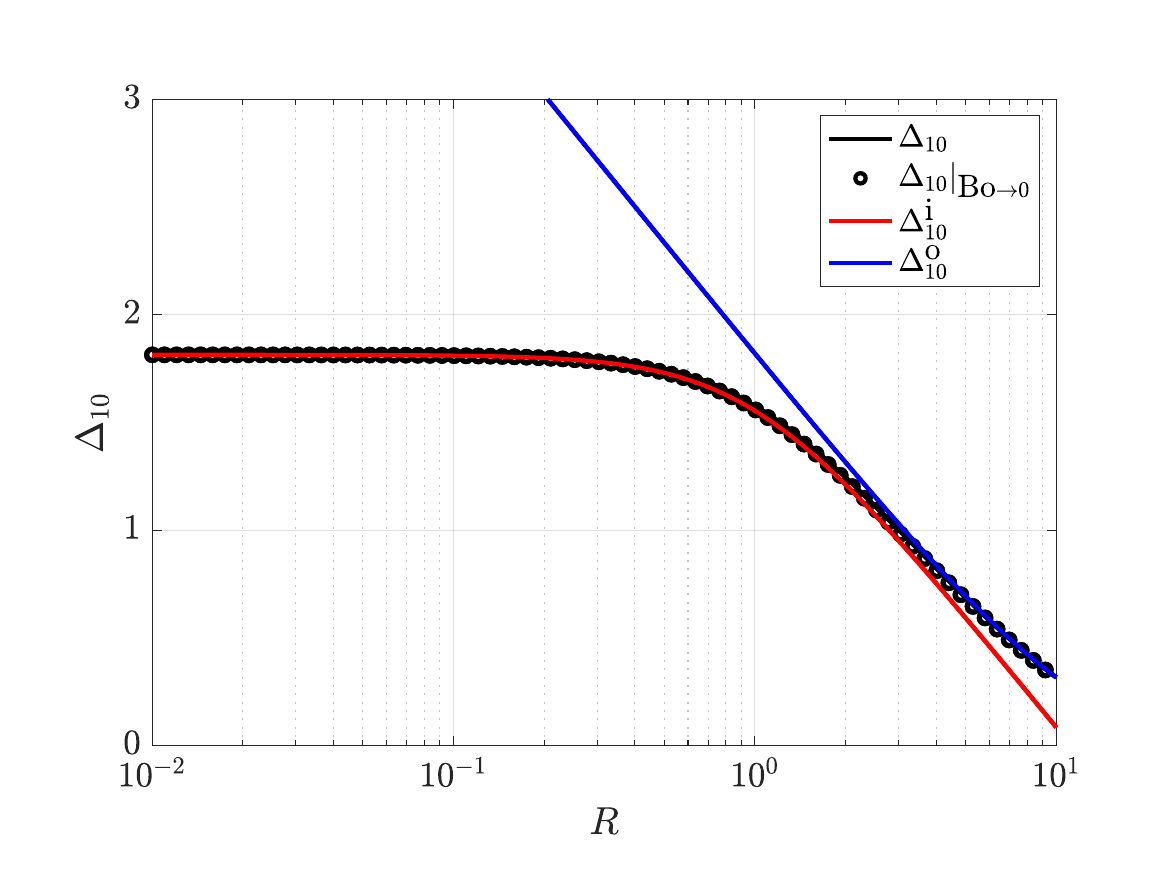}
\caption{Isotropic component $\Delta_{10}$ of the amplitude of the first-order interface deflection as a function of the radial coordinate $R$ (solid black line), as calculated from Eq.~(\ref{eq:delta10_soln}), for $\textrm{Bo} = 0.01$, $D = 1$ and $\dot{D} = 1$. For comparison, the inner solution (red solid line), the outer solution (blue solid line), and the matched crossover expression (symbols), from Eqs.~(\ref{eq:delta_10_inner_soln}),~Eq.~(\ref{eq:delta_10_outer_soln}) and~(\ref{eq:delta_10_inner_soln_matched}) respectively, are also shown.}
\label{fig:axisymmetric_interface_variation}
\end{center}
\end{figure}
The interface deflection is shown in Fig.~\ref{fig:axisymmetric_interface_variation} at a fixed low value of $\textrm{Bo}$. The crossover expression at small $\textrm{Bo}$ described above matches the exact one calculated using Eq.~(\ref{eq:delta10_soln}), with improving precision as $\textrm{Bo}$ reduces. Figure~\ref{fig:axisymmetric_interface_variation} also shows the inner and outer solutions, which diverge in the far and near fields, respectively. 

The other interesting limit arises when $\textrm{Bo}\rightarrow\infty$, leading to the curvature-related terms in Eq.~(\ref{eq:delta_10_diff_eqn}-\ref{eq:delta_11_diff_eqn}) to drop out, giving us: 
\begin{align}
\Delta_{10} = -\frac{P_{00}}{\textrm{Bo}} = \frac{3\dot{D}}{2\textrm{Bo}(D+R^2)^2}~, \\
\Delta_{11} = -\frac{P_{01}}{\textrm{Bo}} = -\frac{6UR}{5\textrm{Bo}(D+R^2)^2}~.
\end{align}
As a consequence, the deflection is directly proportional to the pressure applied with $\kappa/\textrm{Bo} = \eta c l/[g{d^*}^3(\rho_{\textrm{sl}}-\rho)]$ as an effective compliance. From the latter, we see that in the limit of large $\textrm{Bo}$ the surface tension no longer controls the deformation. This large-$\textrm{Bo}$ response is akin to the Winkler response for thin compressible elastic materials, which has been studied previously~\cite{skotheim2005soft,urzay2007elastohydrodynamic,salez2015elastohydrodynamics,Chandler2020,bertin2022soft}. 

\begin{figure}
\begin{subfigure}[b]{0.45\textwidth}
\caption{}
\centering
\includegraphics[width=8cm]{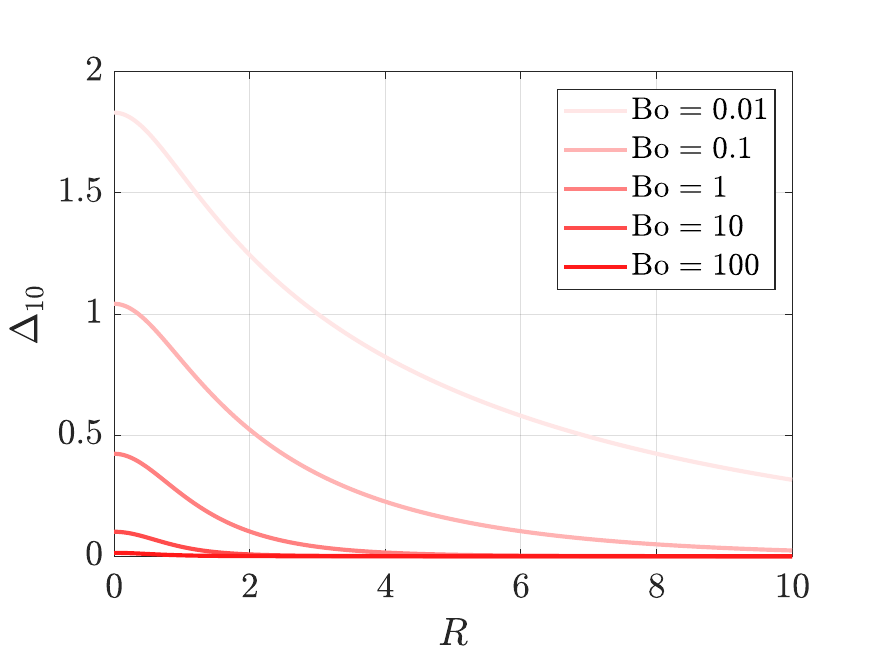}
\label{fig:interface_variation_symmetric_part}
\end{subfigure}
\hfill
\begin{subfigure}[b]{0.45\textwidth}
\caption{}
\centering
\includegraphics[width=8cm]{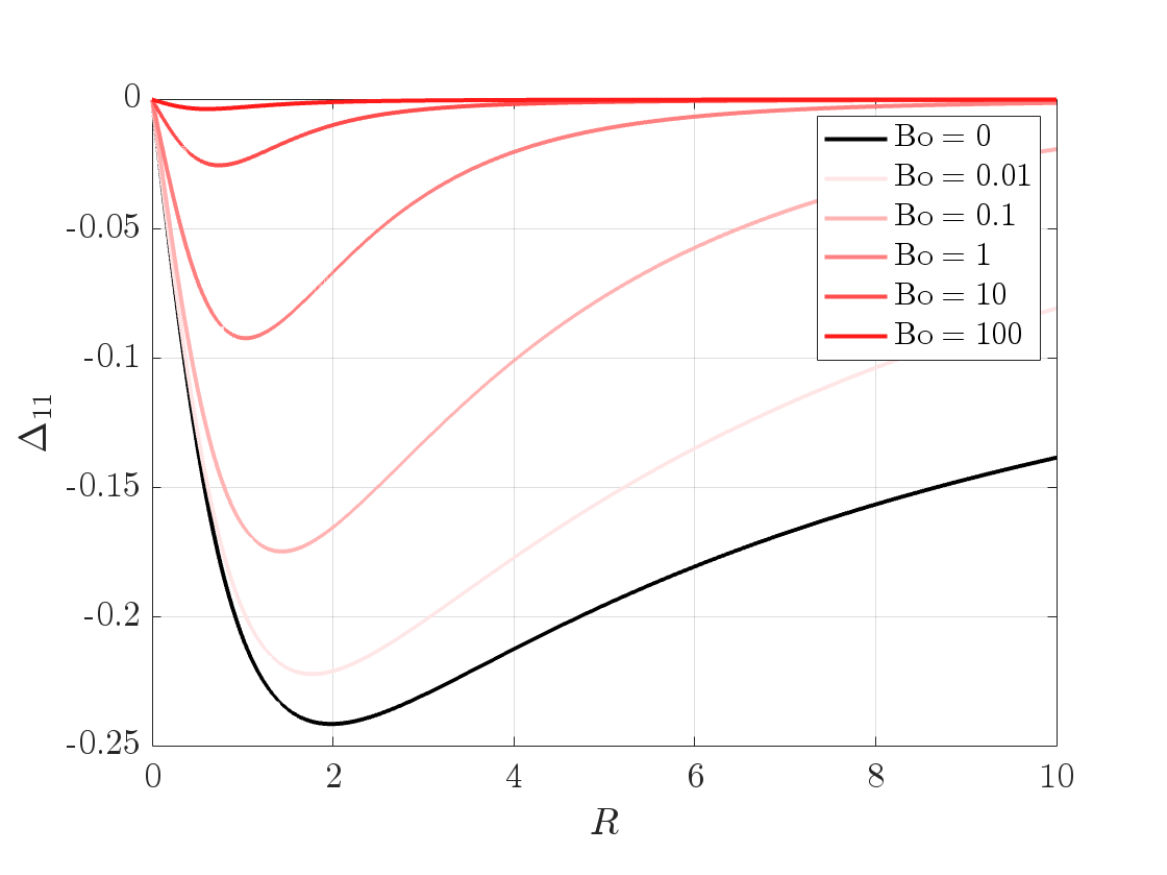}
\label{fig:interface_variation_antisymmetric_part}
\end{subfigure}
\caption{Isotropic component $\Delta_{10}$ (a) and anisotropic component $\Delta_{11}$ (b) of the amplitude of the first-order interface deflection as a function of the radial coordinate $R$, as calculated from Eqs.~(\ref{eq:delta10_soln}) and~(\ref{eq:delta11_soln}), for $D = 1$, various $\textrm{Bo}$ as indicated, and for either $\dot{D} = 1$ (a) or $U=1$ (b). The black solid line denotes the limiting profile for $\textrm{Bo} = 0$ in the anisotropic case, as given in Eq.~(\ref{eq:delta_11_lim}).}
\label{fig:interface_variation}
\end{figure}

Apart from these two extremes, further exploration on the influence of $\textrm{Bo}$ can be done using Eqs.~(\ref{eq:delta10_soln}) and~(\ref{eq:delta11_soln}). Figure~\ref{fig:interface_variation} shows the isotropic and anisotropic profiles of the amplitude of the first-order interface deflection, for different values of $\textrm{Bo}$.  We see that capillarity leads to a long-ranged interface deflection whose range and magnitude both decrease with increasing $\textrm{Bo}$. 

\subsection{First-order pressure}
\begin{figure}
     \begin{subfigure}[b]{0.3\textwidth}
         \caption{}
         \centering
         \includegraphics[width=6cm]{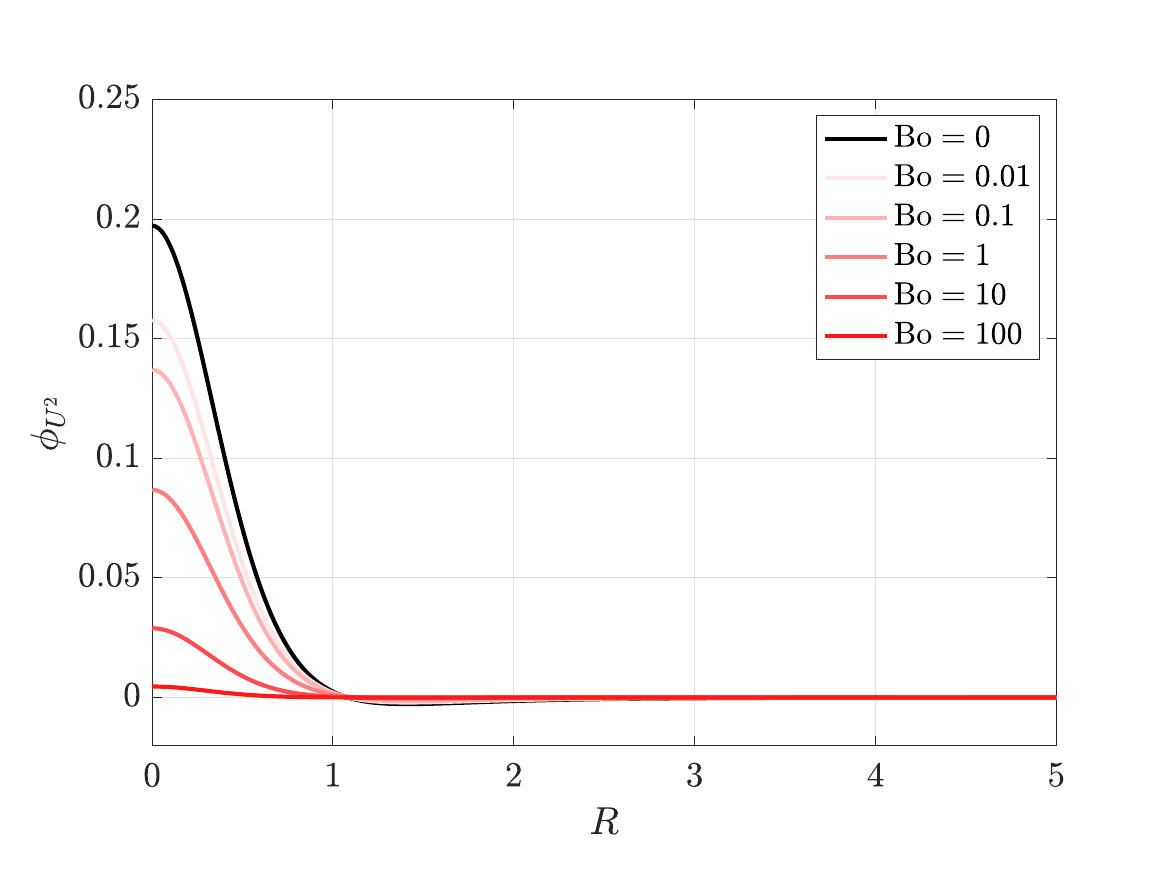}
     \end{subfigure}
     \hfill
     \begin{subfigure}[b]{0.3\textwidth}
     \caption{}
         \centering
         \includegraphics[width=6cm]{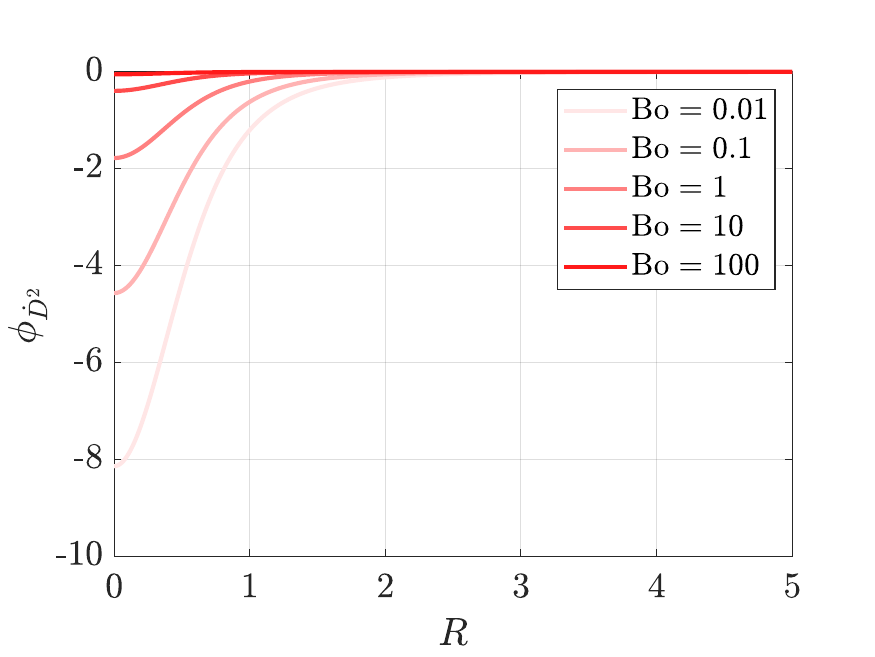}
     \end{subfigure}
     \hfill
     \begin{subfigure}[b]{0.3\textwidth}
     \caption{}
         \centering
         \includegraphics[width=6cm]{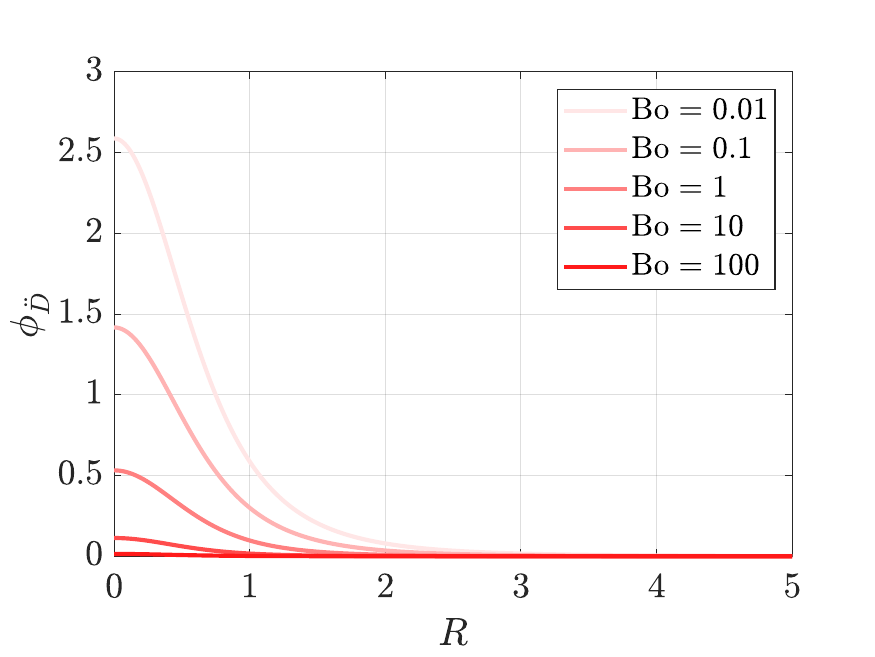}
     \end{subfigure}
    \caption{Auxiliary functions $\phi_k$ (see Eq.~(\ref{eq:p10_decomposition})) of the isotropic 
  component $P_{10}$ of the first-order magnitude $P_1$ of the excess pressure field, as functions of the radial coordinate $R$, for various Bond numbers $\textrm{Bo}$, as obtained by numerically solving Eq.~(\ref{eq:governing_eqn_P_10}) with $D = 1$.}
        \label{fig:p10_auxiliary_functions}
\end{figure}

\begin{figure}
     \begin{subfigure}[b]{0.45\textwidth}
         \caption{}
         \centering
         \includegraphics[width=8cm]{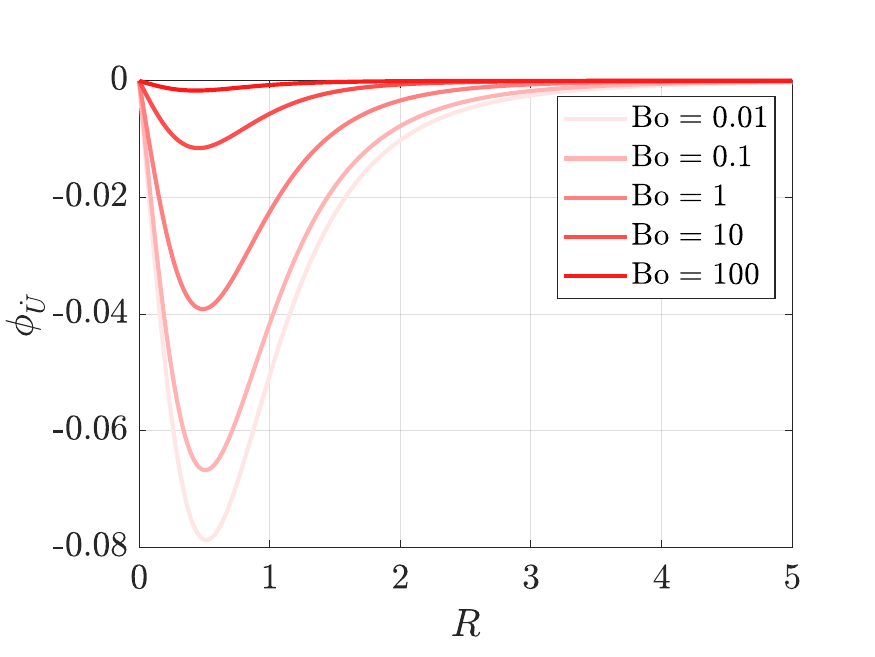}
     \end{subfigure}
     \begin{subfigure}[b]{0.45\textwidth}
     \caption{}
         \centering
         \includegraphics[width=8cm]{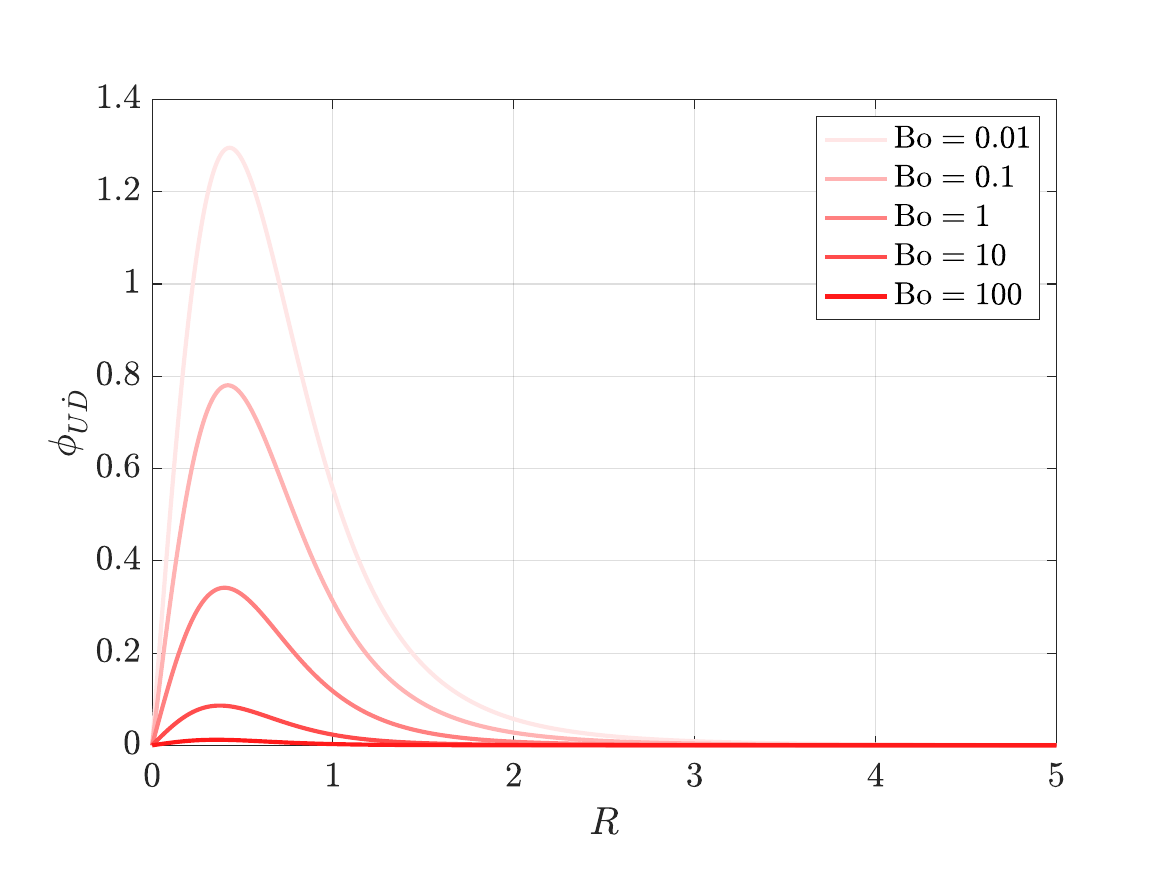}
     \end{subfigure}
    \caption{Auxiliary functions $\phi_k$ (see Eq.~(\ref{eq:p11_decomposition})) of the anisotropic 
  component $P_{11}$ of the first-order magnitude $P_1$ of the excess pressure field, as functions of the radial coordinate $R$, for various Bond numbers $\textrm{Bo}$, as obtained by numerically solving Eq.~(\ref{eq:governing_eqn_P_11}) with $D = 1$.}
        \label{fig:p11_auxiliary_functions}
\end{figure}

At first order in $\kappa$, the pressure field involves in particular the squared velocity, and can thus be decomposed as: 
\begin{align}
P_1(\vec{R},T,\textrm{Bo},D) = P_{10}(R,T,\textrm{Bo},D)+P_{11}(R,T,\textrm{Bo},D)\cos\theta+P_{12}(R,T,\textrm{Bo},D)\cos2\theta~.  
\end{align}
The governing equations for the components $P_{1i}$ of the first-order magnitude $P_1$ of the excess pressure field can be derived by considering Eq.~(\ref{eq:nondim_thin_film}) at first order in $\kappa$. Since $P_{12}$ is not needed to compute the forces and torques, we restrict ourselves to the two following equations:
\begin{align}
\frac{1}{R}\frac{\partial}{\partial R}\left(RH^3\frac{\partial P_{10}}{\partial R}\right) &=\frac{1}{R}\frac{\partial}{\partial R}\left[3RH^2\left(\Delta_{10}\frac{\partial P_{00}}{\partial R}+\frac{\Delta_{11}}{2}\frac{\partial P_{01}}{\partial R}\right)\right]+3U\frac{\partial\Delta_{11}}{\partial R}+3U\frac{\Delta_{11}}{R}-12\frac{\partial \Delta_{10}}{\partial T}~, \label{eq:governing_eqn_P_10}\\
\frac{1}{R}\frac{\partial}{\partial R}\left(RH^3\frac{\partial P_{11}}{\partial R}\right)-\frac{H^3}{R^2}P_{11} &=\frac{1}{R}\frac{\partial}{\partial R}\left[3RH^2\left(\Delta_{10}\frac{\partial P_{01}}{\partial R}+\Delta_{11}\frac{\partial P_{00}}{\partial R}\right)\right]-\frac{3H^2\Delta_{10}P_{01}}{R^2}+6U\frac{\partial\Delta_{10}}{\partial R}-12\frac{\partial \Delta_{11}}{\partial T}~. \label{eq:governing_eqn_P_11}
\end{align}
Using the linearity of the equations above, the components of the pressure field can be expressed as follows: 
\begin{align}
P_{10} &= \frac{U^2}{D^2}\phi_{U^2}+\frac{\dot{D}^2}{D^3}\phi_{\dot{D}^2}+\frac{\ddot{D}}{D^2}\phi_{\ddot{D}}~,\label{eq:p10_decomposition}\\
P_{11} &= \frac{\dot{U}}{D}\phi_{\dot{U}}+\frac{U\dot{D}}{D^2}\phi_{U\dot{D}}~,\label{eq:p11_decomposition}  
\end{align}
where the $\phi_k$ represent the auxiliary functions for the corresponding second-order forcing parameters $k$, such as $U^2$ etc, which all vary with distance $R$ and depend upon the values of $D$ and $\textrm{Bo}$. These functions can then be evaluated by numerically solving Eqs.~(\ref{eq:governing_eqn_P_10}-\ref{eq:governing_eqn_P_11}), together with the boundary conditions that the pressure vanishes in the far field and remains finite and single-valued at the origin. The results are shown in Figs.~\ref{fig:p10_auxiliary_functions} and~\ref{fig:p11_auxiliary_functions}. For all components, as $\textrm{Bo}$ increases, the magnitudes of the auxiliary functions decrease due to the decreasing interface deflection. Interestingly, even though the deflection studied above is long-ranged, the pressure field decays quite rapidly.    

\section{Capillary-lubrication forces and torques}
\begin{figure}[h]
\begin{center}
\includegraphics[width=10cm]{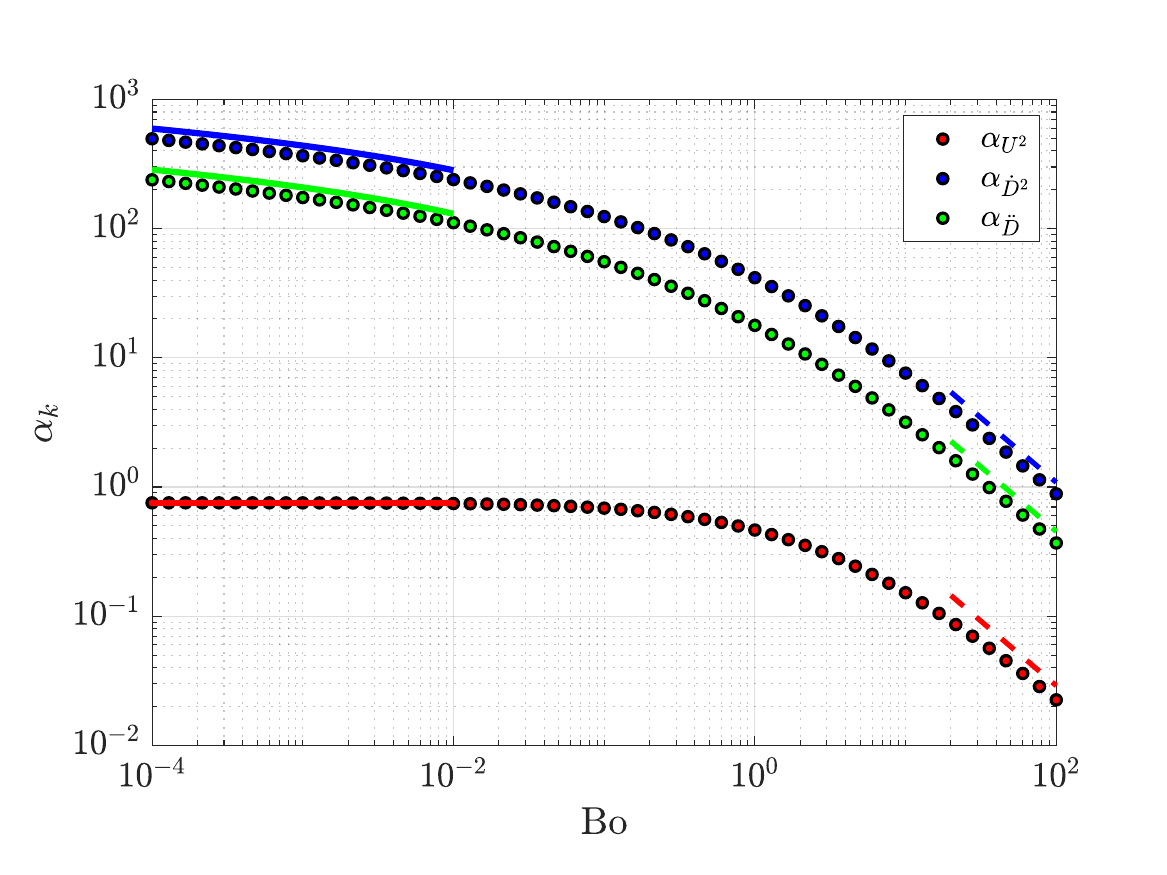}
\caption{Prefactors $\alpha_k$ of the scalings of the vertical-force terms, defined in Eq.~(\ref{eq:normal_force}), as functions of Bond number $\textrm{Bo}$. The solid and dashed lines correspond to the small- and large-$\textrm{Bo}$ behaviors (see Tab.~\ref{tab:coefficients}), respectively}
\label{fig:normal_force}
\end{center}
\end{figure}

Since the zeroth-order forces and torques acting on the sphere are identical to the known ones for the motion near a rigid, flat and no-slip boundary, we focus here on the first-order forces and torques acting on the sphere, and resulting from the interface deflection. These can be calculated from the stress tensor $\mathbf{\Sigma}$, that reads in the lubrication approximation: $\mathbf{\Sigma} \simeq -p\mathbf{I}+\eta\vec{e}_z\partial_z\vec{v}$, where $\mathbf{I}$ denotes the identity tensor. In dimensional units, the first-order vertical force acting on the sphere can be evaluated as:
\begin{align} 
 f_{z,1} \simeq  \frac{2^{5/2}\pi\kappa\eta ca^{3/2}}{{d^*}^{1/2}}\int_0^{\infty} P_{10}(R) R\, \textrm{d}R\ ,    
\end{align}
which can be decomposed using the auxiliary functions calculated before, as: 
\begin{align} 
 f_{z,1} \simeq \alpha_{U^2}\frac{\eta^2 u^2 a^3}{\sigma d^2}-\alpha_{\dot{D}^2}\frac{\eta^2 \dot{d}^2 a^4}{\sigma d^3}+\alpha_{\ddot{D}}\frac{\eta^2 \ddot{d} a^4}{\sigma d^2}\ , \label{eq:normal_force}   
\end{align}
where the $\alpha_k$ (with $k$ indicating here the forcing source, such as $U^2$) are the prefactors of the respective scalings. These prefactors are plotted in Fig.~\ref{fig:normal_force} as functions of $\textrm{Bo}$. An important difference arises between the various prefactors at small $\textrm{Bo}$. Indeed the prefactor $\alpha_{U^2}$ reaches a plateau, whereas $\alpha_{\ddot{D}}$ and $\alpha_{\dot{D}^2}$ increase logarithmically with decreasing $\textrm{Bo}$. These asymptotic behaviors at small $\textrm{Bo}$ can be calculated using Lorentz' reciprocal theorem~\cite{masoud2019reciprocal,Rallabandi2018,daddi2018reciprocal}, by invoking as well Eqs.~(\ref{eq:delta_11_lim}) and~(\ref{eq:delta_10_inner_soln_matched}), as detailed in Appendix and summarised in Tab.~\ref{tab:coefficients}. They are in agreement with the numerical solutions, as shown in Fig.~\ref{fig:normal_force}. Besides, as $\textrm{Bo}$ increases, all the prefactors decrease, and this decrease becomes inversely proportional to $\textrm{Bo}$ in the large-$\textrm{Bo}$ limit, highlighting the transition to the Winkler-like regime. The corresponding asymptotic expressions have been calculated previously~\cite{bertin2022soft}, are summarised in Tab.~\ref{tab:coefficients}, and are in agreement with the numerical solutions, as shown in Fig.~\ref{fig:normal_force}. 

\begin{figure}[h]
\begin{center}
\includegraphics[width=10cm]{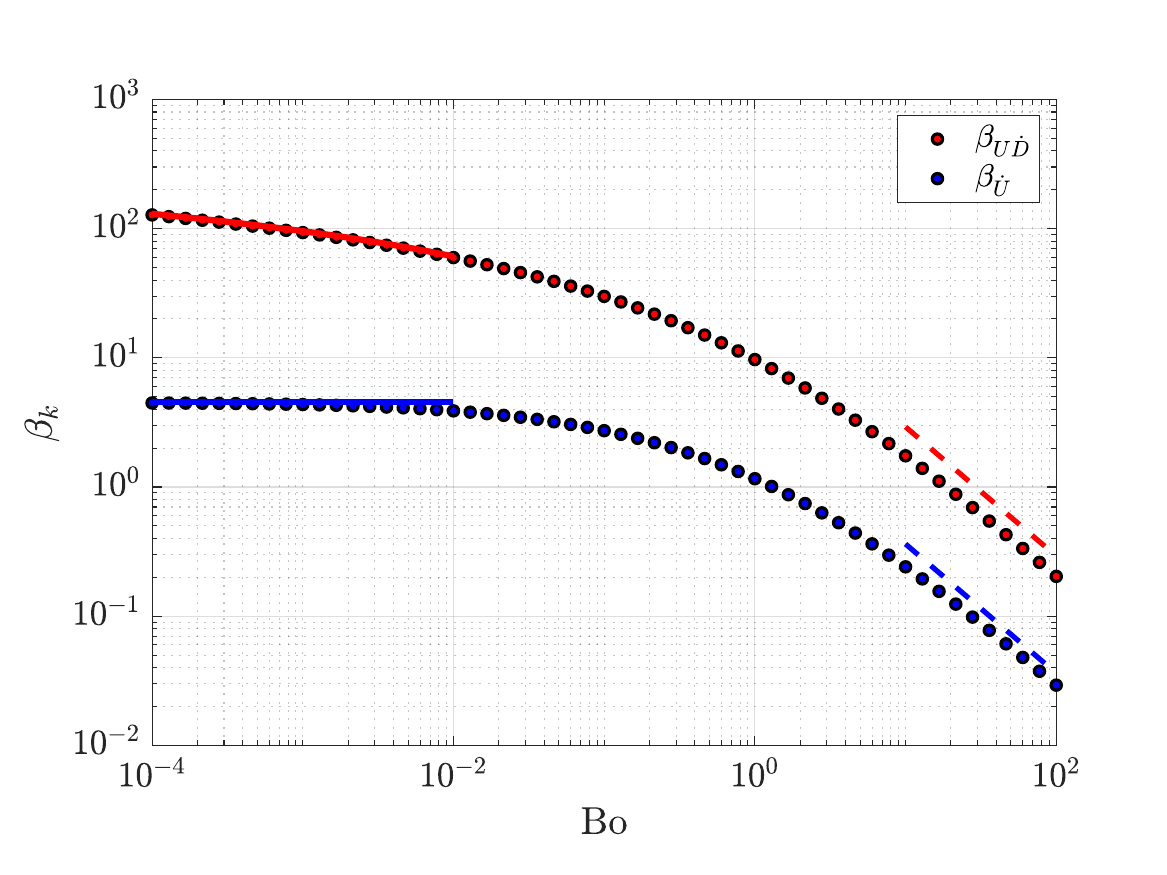}
\caption{Prefactors $\beta_k$ of the scalings of the horizontal-force terms, defined in Eq.~(\ref{eq:drag_force}), as functions of Bond number $\textrm{Bo}$. The solid and dashed lines correspond to the small- and large-$\textrm{Bo}$ behaviors (see Tab.~\ref{tab:coefficients}), respectively.}
\label{fig:drag_force}
\end{center}
\end{figure}

Similarly, the horizontal force acting on the sphere is given by the expression~\cite{bertin2022soft}:  
\begin{multline} 
 f_{x,1} \simeq 2\pi\eta c a\kappa\int_0^\infty \left[-2RP_{11}-\frac{H}{2}\left(\partial_R P_{11}+\frac{P_{11}}{R}\right)+\frac{\Delta_{11}}{2}\partial_R P_{00}+\frac{\Delta_{10}}{2}\left(\partial_R P_{01}+\frac{P_{01}}{R}\right)-2\frac{U\Delta_{10}}{H^2}\right]R\textrm{d}R\ , 
 \end{multline}
 which can be decomposed into:
 \begin{align} 
f_{x,1} \simeq -\beta_{U\dot{D}}\frac{\eta^2 u \dot{d}a^3}{\sigma d^2}+\beta_{\dot{U}}\frac{\eta^2  \dot{u}a^3}{\sigma d}~,\label{eq:drag_force}   
 \end{align}
where the $\beta_k$ (with $k$ indicating here the forcing source, such as $\dot{U}$) are the prefactors of the respective scalings. These prefactors are plotted in Fig.~\ref{fig:drag_force} as functions of $\textrm{Bo}$. An important difference arises once again between the two prefactors at small $\textrm{Bo}$. Indeed the prefactor $\beta_{\dot{U}}$ reaches a plateau, whereas $\beta_{U\dot{D}}$ increases logarithmically with decreasing $\textrm{Bo}$. These asymptotic behaviors at small $\textrm{Bo}$ can be once again calculated using Lorentz' reciprocal theorem, as detailed in Appendix and summarised in Tab.~\ref{tab:coefficients}. They are in agreement with the numerical solutions, as shown in Fig.~\ref{fig:drag_force}. Besides, as $\textrm{Bo}$ increases, all the prefactors decrease, and this decrease becomes inversely proportional to $\textrm{Bo}$ in the large-$\textrm{Bo}$ limit, highlighting once again the transition to the Winkler-like regime. The corresponding asymptotic expressions have been calculated previously~\cite{bertin2022soft}, are summarised in Tab.~\ref{tab:coefficients}, and are in agreement with the numerical solutions, as shown in Fig.~\ref{fig:drag_force}. 

As shown in previous studies~\cite{salez2015elastohydrodynamics,bertin2022soft}, the contributions to the first-order torque felt by the sphere along the $y$-axis have the same numerical prefactors as those for the first-order horizontal force, with the inclusion of a supplementary length-scale factor $a$. Hence, the first-order torque exerted on the sphere is given by: 
\begin{align} 
\tau_{y,1} \simeq \beta_{U\dot{D}}\frac{\eta^2 u \dot{d}a^4}{\sigma d^2}-\beta_{\dot{U}}\frac{\eta^2  \dot{u}a^4}{\sigma d}~.\label{eq:torque}   
\end{align}

We conclude this whole section by an important remark. The crossover from the capillary-dominated to the Winkler-like regime occurs at $\textrm{Bo} \sim 1$. This occurs when the hydrodynamic radius is comparable to the capillary length. Since typical capillary lengths are on the order of $\sim1$~mm, and given the lubrication conditions, such a crossover can only be felt with spheres of millimetric radii and above.
\renewcommand{\arraystretch}{2}
\begin{table}[]
    \centering
\begin{tabular}{|| c | c | c||} 
 \hline
 $\alpha_k\, ,\,\beta_k$ & $\textrm{Bo}\rightarrow0$ & $\textrm{Bo}\rightarrow\infty$ \\[1ex] 
 \hline\hline
 $\alpha_{U^2}$ & $6\pi/25$ & $96\pi/(125\textrm{Bo})$  \\ 
 \hline
 $\alpha_{\dot{D}^2}$ & $-6\pi[2+6\gamma+3\ln(\textrm{Bo}/4)]$ & $144\pi/(5\textrm{Bo})$  \\ 
 \hline
 $\alpha_{\ddot{D}}$ & $-9\pi[1+2\gamma+\ln(\textrm{Bo}/4)]$ & $12\pi/\textrm{Bo}$  \\ 
 \hline
 $\beta_{U\dot{D}}$ & $-24\pi[4+10\gamma+5\ln(\textrm{Bo}/4)]/25$ & $968\pi/(125\textrm{Bo})$  \\ 
 \hline
 $\beta_{\dot{U}}$ & $36\pi/25$ & $24\pi/(25\textrm{Bo}) $ \\  [1ex] 
 \hline
\end{tabular}
    \caption{Asymptotic behaviours of the scaling prefactors $\alpha_k$ and $\beta_k$ at small (see Appendix) and large (see~\cite{bertin2022soft}) Bond numbers $\textrm{Bo}$, where $\gamma$ denotes Euler's constant. }
    \label{tab:coefficients}
\end{table}

\section*{Conclusion}
Using capillary-lubrication theory, scaling arguments and numerical integrations, we explored the asymptotic forces and torques generated on a sphere in translational motion within a viscous fluid, in close proximity to a deformable capillary interface with another, infinitely viscous fluid on the other side. Due to a symmetry between translational and rotational motions in soft lubrication~\cite{bertin2022soft}, our work immediately generalizes to the case where rotation  would be added. Specifically, by employing a perturbation analysis in the limit of small deformation of the fluid interface, we calculated the pressure fields decomposed into their various contributions from different degrees of freedom of the sphere. We investigated in particular the effects of gravity, which not only changes the scaling laws of the forces and torques, but also show a Winkler-like elastic response at large Bond numbers. Altogether, our results allow to quantify and possibly control soft-lubricated motion near tensile interfaces, which are ubiquitous in soft matter and biological physics. For example, the viscocapillary lift force revealed among others by our analysis, could contribute to rationalize the swimming of active organisms at the air-water interface~\cite{trouilloud2008soft}, the self-propulsion of levitating Leidenfrost droplets~\cite{gauthier2019self} for which a detailed mechanism is lacking to date, or the thermal motion of oil droplets near rigid walls for which intriguing transient non-conservative forces have been reported~\cite{fares2024observation}.

\begin{acknowledgments}
The authors thank Ga\"elle Aud\'eoud, Vincent Bertin and Isabelle Cantat for interesting discussions, as well as Nicolas Fares for noticing a typographical mistake in Eq.~(4.2) of the publisher's version (\textit{i.e.} Eq.~(42) here). They acknowledge financial support from the European Union through the European Research Council under EMetBrown (ERC-CoG-101039103) grant. Views and opinions expressed are however those of the authors only and do not necessarily reflect those of the European Union or the European Research Council. Neither the European Union nor the granting authority can be held responsible for them. The authors also acknowledge financial support from the Agence Nationale de la Recherche under Softer (ANR-21-CE06-0029) and Fricolas (ANR-21-CE06-0039) grants. Finally, they thank the Soft Matter Collaborative Research Unit, Frontier Research Center for Advanced Material and Life Science, Faculty of Advanced Life Science at Hokkaido University, Sapporo, Japan.  
\end{acknowledgments}

\section*{Declaration of Interests}
The authors report no conflict of interest. 

\section*{Appendix: Lorentz reciprocal theorem}
In this appendix, we employ Lorentz reciprocal theorem for Stokes flows~\cite{Rallabandi2018,masoud2019reciprocal,daddi2018reciprocal} in the limit of vanishing Bond number $\textrm{Bo}$, which allows us to evaluate the asymptotic behaviors of the scaling prefactors $\alpha_k$ and $\beta_k$.

\subsection*{A: Vertical force}
The model problem introduced to perform the calculation comprises a sphere moving in a viscous fluid and towards an immobile, rigid, planar surface. We note $\vec{\hat{V}}_{\perp} = -\hat{V}_{\perp}\vec{e}_z$, the velocity at the particle surface while assuming a no-slip boundary condition at the wall surface located at $z=0$. The model problem is described by the incompressible Stokes' equations, $\nabla\cdot\mathbf{\hat{\Sigma}}=0$ and $\nabla\cdot\vec{\hat{v}}_{\perp}=0$, where $\mathbf{\hat{\Sigma}}$ denotes the stress tensor of the model problem and $\vec{v}_{\perp}$ denotes the corresponding fluid velocity. We invoke lubrication theory to obtain the corresponding pressure and velocity fields, given by: 
\begin{align}
    \hat{p}_{\perp}(\vec{r}) &= \frac{3\eta\hat{V}_{\perp}a}{\hat{h}^2(r)}~, \label{eq:AppendixA:model1_pressure}\\
    \mathbf{\hat{v}}_{\perp}(\vec{r},z) &= \frac{\nabla \hat{p}_{\perp}(\vec{r})}{2\eta}z[z-\hat{h}(r)]~, \label{eq:AppendixA:model2_velocity}
\end{align}
where:
\begin{align}
        \hat{h}(r) &\simeq d+\frac{r^2}{2a}~. \label{eq:AppendixA:model_height}
\end{align}

From the Lorentz reciprocal theorem, one has:
\begin{align}
    \int_{\mathcal{S}}\vec{n}\cdot\mathbf{\Sigma}\cdot\vec{\hat{v}}_{\perp}\,\textrm{d}s = \int_{\mathcal{S}}\vec{n}\cdot\mathbf{\hat{\Sigma}}\cdot\vec{v}\,\textrm{d}s~,
\end{align}
where $\mathbf{\Sigma}$ and $\vec{v}$ denote the stress tensor and flow velocity for the real problem, with $\mathcal{S}$ denoting the entire bounding surface, including the surface of the sphere, the surface of the substrate and the surface located at $\vec{r}\rightarrow\infty$. The unit vector normal to the surface pointing towards the fluid is denoted by $\mathbf{n}$. Given the boundary conditions of the model problem, the above relation simplifies to give:
\begin{align}
    F_z = -\frac{1}{\hat{V}_{\perp}}\int_{\mathcal{S}}\vec{n}\cdot\mathbf{\hat{\Sigma}}\cdot\vec{v} \textrm{d}s~.
\end{align}
To approximate the velocity field at the wall surface, we perform a Taylor expansion accounting for the small deformation of the wall, as:
\begin{align}
    \vec{v}|_{z=0} &\simeq \vec{v}|_{z=\delta}-\delta\partial_z\vec{v}_0|_{z=0}~,\\
    &= -u\vec{e}_x-\dot{d}\vec{e}_z+(\partial_t-u\partial_x)\delta\vec{e}_z-\delta\partial_z\vec{v}_0|_{z=0}~, 
\end{align}
where $\vec{v}_0$ denotes the zeroth-order velocity field corresponding to a sphere moving near a rigid surface. Thus, the leading-order normal force simplifies into:
\begin{align}
    F_{z,1} \simeq -\frac{1}{\hat{V}_{\perp}}\int_{\mathbb{R}^2}(\hat{p}_{\perp}(\partial_t-u\partial_x)\delta+\eta\partial_z\vec{\hat{v}}_{\perp}|_{z=0}\cdot\partial_z\vec{v}_0|_{z=0})\,\textrm{d}\vec{r}~.
\end{align}
Computing the latter integral by considering the deflection $\delta$ (or $\Delta$ in dimensionless variables) generated at vanishing $\textrm{Bo}$ allows us to recover the asymptotic expressions presented in Tab.~\ref{tab:coefficients}. 

\subsection*{B: Horizontal force}
The model problem consists here of a sphere translating with a velocity $\hat{V}_{\parallel}\vec{e}_x$, parallel and near an immobile and rigid substrate with no-slip boundary conditions applied at both the surfaces of the sphere and the substrate. A similar treatment as in the previous section leads to:
\begin{align}
    \hat{p}_{\parallel}(\vec{r}) &= \frac{6\eta\hat{V}_{\parallel}r\cos\theta}{5\hat{h}^2(r)}~, \label{eq:AppendixB:model2_pressure}\\
    \vec{\hat{v}}_{\parallel}(\vec{r},z) &= \frac{\nabla \hat{p}_{\parallel}(\vec{r})}{2\eta}z[z-\hat{h}(r)]+\vec{\hat{V}}_{\parallel}\frac{z}{\hat{h}(r)}~,  \label{eq:AppendixB:model2_velocity}
\end{align}
and:
\begin{align}
    F_{x,1} \simeq -\frac{1}{\hat{V}_{\parallel}}\int_{\mathbb{R}^2}(\hat{p}_{\parallel}(\partial_t-u\partial_x)\delta+\eta\partial_z\vec{\hat{v}}_{\parallel}|_{z=0}\cdot\partial_z\vec{v}_0|_{z=0})\textrm{d}\vec{r}~.
\end{align}
Computing the latter integral by considering the deflection $\delta$ (or $\Delta$ in dimensionless variables) generated at vanishing $\textrm{Bo}$ allows us to recover the asymptotic expressions presented in Tab.~\ref{tab:coefficients}. 
\bibliography{Jha2024}
\end{document}